# Permutation Decoding and the Stopping Redundancy Hierarchy of Linear Block Codes

Thorsten Hehn*, Olgica Milenkovic†, Stefan Laendner†, Johannes B. Huber*
*Institute for Information Transmission, University of Erlangen-Nuremberg, Germany
†Department of Electrical and Computer Engineering, University of Colorado, Boulder
*{hehn, huber}@LNT.de, †{laendner, milenkov}@colorado.edu

*Abstract*— We investigate the stopping redundancy hierarchy of linear block codes and its connection to permutation decoding techniques. An element in the ordered list of stopping redundancy values represents the smallest number of possibly linearly dependent rows in any parity-check matrix of a code that avoids stopping sets of a given size. Redundant parity-check equations can be shown to have a similar effect on decoding performance as permuting the coordinates of the received codeword according to a selected set of automorphisms of the code. Based on this finding we develop new decoding strategies for data transmission over the binary erasure channel that combine iterative message passing and permutation decoding in order to avoid errors confined to stopping sets. We also introduce the notion of s-SAD sets, containing the smallest number of automorphisms of a code with the property that they move any set of not more than $s$ erasures into positions that do not correspond to stopping sets within a judiciously chosen parity-check matrix.

## I. INTRODUCTION

Stopping sets are configurations of variable nodes in Tanner graphs of codes on which iterative decoders designed for the binary erasure channel (BEC) fail [1]. The size and number of stopping sets depends on the particular choice of the parity-check matrix used for decoding. Since a large number of rows in the parity-check matrix of a code ensures increased flexibility in terms of meeting predefined constraints on the structure of stopping sets, several authors recently proposed using redundant parity-check matrices to improve the performance of iterative decoders [2], [3], [4], [5]. Adding judiciously chosen redundant rows into a parity-check matrix improves the performance of the code but also increases the overall complexity of decoding. This motivates the study of possible trade-offs between the number of redundant rows and the size of the smallest stopping set in a parity-check matrix of the code. In this context, Schwartz and Vardy [2] introduced the *stopping redundancy* of an $[n, k, d]$ linear block code $\mathcal{C}$ to represent the smallest number of codewords that span the dual $\mathcal{C}^\perp$ code and constitute a matrix with no stopping sets of size smaller than $d$. The same authors also provided lower and upper bounds on the stopping redundancy, the latter growing exponentially with the co-dimension of the code for most examples considered. This finding raised the question if there exist codes for which one could significantly decrease the number of redundant rows in the parity-check matrix by slightly decreasing the size of its smallest stopping set. Several results regarding this problem were derived by Abdel-Ghaffar and Weber [3], Hehn et. al. [5], as well as by Hollman and Tolhuizen [6], who addressed the question of determining the smallest redundancy of a parity-check matrix of a code that allows for decoding all *correctable erasure patterns*.

We address the problem of improving the decoding performance of codes by adding redundant parity-check equations from a different perspective. Our contributions are two-fold: the first contribution consists in introducing the notion of the *stopping redundancy hierarchy* of a linear block code; the second contribution consists in demonstrating how the stopping redundancy hierarchy can be studied in the context of *permutation decoding*, first proposed for decoding of cyclic codes over the binary symmetric channel in [7]. In the latter context, we describe a new decoding strategy that represents a combination of iterative message passing and permutation decoding. Such an algorithm operates on properly designed *non-redundant* parity-check matrices, and it has the property that it sequentially moves collections of erasures confined to stopping sets into positions that do not correspond to stopping sets. Furthermore, we introduce a new class of code invariants related to the structure of their automorphism group, termed **s-SAD** ($s$-Stopping Automorphism Group Decoding) sets.

The paper is organized as follows. Section II introduces relevant definitions and terminology, including the notion of the *stopping redundancy hierarchy* and *permutation decoding*. Section III provides an overview of some straightforward analytical results regarding **s-SAD** sets. The performance of the new iterative decoding strategy that uses knowledge about the automorphism group of the code is described in Section IV.

## II. DEFINITIONS AND TERMINOLOGY

Henceforth, we focus our attention on binary linear $[n, k, d]$ block codes $\mathcal{C}$ used for signalling over the *binary erasure channel* with erasure rate $0 < \text{ER} < 1$. For decoding, an iterative belief-propagation algorithm is used on a suitable chosen parity-check matrix $\boldsymbol{H}$ of $\mathcal{C}$. It can be shown that for this combination of transmission channel and message recovery procedure, all decoding errors are confined to collections of variable nodes containing *stopping sets* [1]. Stopping sets can be formally defined as follows. Let the columns of $\boldsymbol{H}$ be indexed by $J = \{0, \ldots, n-1\}$. For a set $I \in J$, the *restriction* of $\boldsymbol{H}$ to $I$ is the set of columns of $\boldsymbol{H}$ indexed by $I$. Here, it is tacitly assumed that rows of zero weight in the restriction are removed from the underlying array.

*Definition 2.1:* For a given parity-check matrix $\boldsymbol{H}$ of $\mathcal{C}$, a *stopping set* $\mathcal{S}_\sigma(n)(\boldsymbol{H})$ of size $\sigma$ is a set of $\sigma$ columns for which the restriction of $\boldsymbol{H}$ does not contain rows of Hamming weight one. The *stopping distance* of a parity-check matrix $\boldsymbol{H}$ is the size of the smallest stopping set in $\boldsymbol{H}$.

Note that the size and the number of stopping sets in the decoder graph depends on the particular choice of the parity-check matrix. It is straightforward to see that adding rows to a fixed parity-check matrix $\boldsymbol{H}$ may only increase its stopping distance. The added rows represent linear combinations of the bases vectors in $\boldsymbol{H}$, and are referred to as *redundant rows (redundant parity-checks)*. A parity-check matrix $\boldsymbol{H}$ containing redundant parity-check equations is henceforth referred to as a *redundant parity-check matrix*. The term parity-check matrix is consequently reserved for matrices of smallest possible dimension, i.e. for matrices with dimension $(n-k) \times n$.

In order to quantify the relationship between the size of the smallest stopping set in a redundant parity-check matrix and its number of rows, we study the *stopping redundancy hierarchy* of a code. The related notion of the *stopping redundancy* of a code was first described in [2], while the latter set of parameters was introduced by the authors in [5].

*Definition 2.2:* Let $\mathcal{C}$ be a linear code with minimum distance $d$. For $\ell \leq d$, the $\ell$-th stopping redundancy of $\mathcal{C}$ is the smallest integer $\rho_\ell(\mathcal{C})$ such that there exists a (possibly redundant) parity-check matrix $\boldsymbol{H}$ of $\mathcal{C}$ with $\rho_\ell(\mathcal{C})$ rows and stopping distance at least $\ell$. The ordered set of integers

$$(\rho_2(\mathcal{C}), \rho_3(\mathcal{C}), \ldots, \rho_d(\mathcal{C}))$$

is called the *stopping redundancy hierarchy* of $\mathcal{C}$. The integer $\rho_d(\mathcal{C})$ is the stopping redundancy of $\mathcal{C}$.

*Example 2.1:* Consider the class of cyclic and extended cyclic codes studied in [5]. There, it was shown that redundant cyclic parity-check matrices, consisting of $m$ cyclic shifts of a generator codeword of the dual code, $n-k \leq m \leq n$, have excellent stopping distance properties. As an illustration, such matrices were used to find the following constructive bounds on the stopping redundancy hierarchy of the Golay $[23, 11, 7]$ code $\mathcal{G}_{23}$

$$\rho_4(\mathcal{G}_{23}) = 11, \rho_5(\mathcal{G}_{23}) \leq 15, \rho_6(\mathcal{G}_{23}) \leq 18, \rho_7(\mathcal{G}_{23}) \leq 23. \quad (1)$$

Let us now take a closer look at the properties of decoders operating on cyclic redundant parity-check matrices. Using a collection of cyclically shifted rows in $\boldsymbol{H}$ has the same effect as using only *one of these rows* and *cyclically shifting the received codeword*. In general, the same claim is true when one uses a collection of redundant rows that represent the images of one given row in the parity-check matrix under a set of coordinate permutations. Such coordinate permutations must necessarily preserve the codeword structure, i.e. they must correspond to automorphisms of the code. Furthermore, such automorphisms have to be chosen in a way that they relocate the positions of the erasures in the received codeword to coordinates that do not correspond to stopping sets in the parity-check matrix of the code.

To put the described observations in a rigorous mathematical framework, we first recall the definition of the *automorphism group* of a code.

*Definition 2.3:* [7] The permutations which send $\mathcal{C}$ into itself, i.e. codewords go into (possibly different) codewords, form the automorphism group of $\mathcal{C}$, denoted by $\text{Aut}(\mathcal{C})$. If $\mathcal{C}$ is a linear code and $\mathcal{C}^\perp$ is its dual code, then $\text{Aut}(\mathcal{C}) = \text{Aut}(\mathcal{C}^\perp)$.

Decoding procedures that use information about the automorphism group of a code have a long history [7]. Such procedures are known as *permutation decoding algorithms* and they were traditionally restricted to decoding of messages transmitted over binary symmetric channels (BSCs)[1]. Within the framework of permutation decoding, a codeword $\boldsymbol{c} \in \mathcal{C}$, corrupted by an error vector $\boldsymbol{e} = [e_0, e_1, \ldots, e_{n-1}]$ of weight less than or equal to $t$, $2t + 1 \leq d$, is subject to the following procedure. First, a parity-check matrix of the code of the form $[\boldsymbol{A}|\boldsymbol{I}]$, where $\boldsymbol{I}$ is the identity matrix of order $n-k$ is chosen. Then, the syndrome of the received vector $\boldsymbol{y} = \boldsymbol{c} + \boldsymbol{e}$, $\boldsymbol{z} = \boldsymbol{H}\boldsymbol{y}^\text{T}$, is computed. If the weight of $\boldsymbol{z}$ is greater than $t$, the vector $\boldsymbol{y}$ is permuted according to a randomly chosen automorphism. This process is repeated until either all automorphisms are tested or until the syndrome has weight less than or equal to $t$. In the former case, the decoder declares an error. In the latter case, all decoding errors are provably confined to parity-check positions so that decoding terminates by recovering the information symbols.

For the purpose of permutation decoding, one would like to identify the smallest set of automorphisms that moves any set of $h \leq t$ positions in $[0, n-1]$ into the parity-check positions $k, k+1, \ldots, n-1$ of $[\boldsymbol{A}|\boldsymbol{I}]$.

*Definition 2.4:* [9] If $\mathcal{C}$ is a $t$-error correcting code with information set $\mathcal{I}$ and parity-check set $\mathcal{P}$, then a **PD**($\mathcal{C}$)-set (permutation decoding set of $\mathcal{C}$) is a set $S$ of automorphisms of $\mathcal{C}$ such that every $t$-set of coordinate positions is moved by at least one member of $S$ into the check-positions $\mathcal{P}$. For $s \leq t$, an **s-PD**($\mathcal{C}$)-set is a set $S$ of automorphisms of $\mathcal{C}$ such that every $s$-set of coordinate positions is moved by at least one member of $S$ into $\mathcal{P}$.

Throughout the remainder of the paper, we will be concerned with **PD** sets of *smallest possible size*, and we simply refer to them as **PD** sets. Clearly, **PD** and **s-PD** sets may not exist for a given code and complete or partial information about **PD** sets is known in very few cases [9]. Nevertheless, even this partial information can be used to derive useful results regarding the analogues of **PD** sets for iterative decoders operating on stopping sets.

In the next section, we study the connection between the automorphism group of a code and its stopping redundancy hierarchy. In order to distinguish between iterative decoders that use automorphisms to reduce errors due to stopping sets and standard permutation decoders, we refer to the former as *automorphism group decoders*. Automorphism group decoders offer one significant advantage over iterative decoders operating on redundant parity-check matrices: they have very low storage complexity and at the same time excellent decoding performance. This is, to a certain degree, offset by the slightly increased computational complexity of automorphism group decoders.

---

[1] Recently, permutation decoders were also used for decoding of messages transmitted over the AWGN channel [8], although not for the purpose of finding error-free information sets nor for the purpose of eliminating pseudocodewords such as stopping sets.

### III. STOPPING REDUNDANCY, PD, AND SAD SETS

We provide next a sampling of results regarding a generalization of the notion of **PD** sets termed *stopping automorphism group decoding* (**SAD**) sets. We then proceed to relate **SAD** sets to the stopping redundancy hierarchy of a code.

*Definition 3.1:* Let $H$ be a parity-check matrix of an error-correcting code $\mathcal{C}$ with minimum distance $d$. A **SAD**($H$) set of $H$ is the smallest set $S$ of automorphisms of $\mathcal{C}$ such that every $b$-set of coordinate positions, $1 \leq b \leq d-1$, is moved by at least one member of $S$ into positions that do not correspond to a stopping set of $H$. Similarly, for $s \leq d-1$, an **s-SAD**($H$)-set is the smallest set of automorphisms of $\mathcal{C}$ such that every $b$-set of coordinate positions, $b \leq s$, is moved by at least one member of $S$ into positions that do not correspond to a stopping set in $H$. For a given code $\mathcal{C}$, we also define

$$\mathbf{S}_s^\star(\mathcal{C}) = \min_{H(\mathcal{C})} |\mathbf{s}\text{-}\mathbf{SAD}(H(\mathcal{C}))|,$$
$$\mathbf{S}^\star(\mathcal{C}) = \min_{H(\mathcal{C})} |\mathbf{SAD}(H(\mathcal{C}))|,$$

and refer to $\mathbf{S}_s^\star(\mathcal{C})$ and $\mathbf{S}^\star(\mathcal{C})$ as to the *s-automorphism redundancy* and *automorphism redundancy* of $\mathcal{C}$.

For a given code, **s-SAD** sets may not exist. This is a consequence of the fact that there may be no automorphisms that move all arbitrary collections of not more than $s$ coordinates into positions that do not correspond to a stopping set in any given parity-check matrix. But whenever such sets exist, they can be related to the stopping redundancy hierarchy and **PD** sets of the code. First, it is straightforward to show that for all $1 \leq s \leq \lfloor \frac{d-1}{2} \rfloor$ one has $\mathbf{S}_s^\star(\mathcal{C}) \leq |\mathbf{s}\text{-}\mathbf{PD}(\mathcal{C})|$, whenever such sets exist. This follows from considering parity-check matrices in systematic form, and from the Singleton bound, which asserts that for any code, one must have $d-1 \leq n-k$. Furthermore, it is straightforward to see that for a restricted set of parity-check matrices, automorphism group decoders trade redundant rows with automorphisms. This is formally described by the following lemma, the proof of which is straightforward and hence omitted.

*Lemma 3.1:* Let $\mathcal{C}$ be an $[n,k,d]$ code. Then

$$\rho_s(\mathcal{C}) \leq (n-k) \times \mathbf{S}_s^\star(\mathcal{C}), \tag{2}$$

for all $1 \leq s \leq \lfloor \frac{d-1}{2} \rfloor$, provided that an **s-SAD**($\mathcal{C}$) set exists.

One class of codes for which it is straightforward to prove the existence of certain **SAD** sets is the class of codes with *transitive* automorphism groups, described below.

*Definition 3.2:* A group $G$ of permutations of the symbols $[0, n-1] = \{1, 2, \ldots, n-1\}$ is transitive if for any two symbols $i, j$ there exists a permutation $\pi \in G$ such that $i\pi = j$. A group is said to be $t$-fold transitive if for any two collections of distinct numbers $i_1, \ldots, i_t \in [0, n-1]$ and $j_1, \ldots, j_t \in [0, n-1]$, there exists a $\pi \in G$ such that $i_1\pi = j_1, \ldots, i_t\pi = j_t$.

*Lemma 3.2:* Let $\mathcal{C}$ be a code with an $s$-transitive automorphism group. Then there exist **b-SAD** sets of $\mathcal{C}$ for all $b \leq s$.

*Proof:* Let $H$ be of the form $[A|I]$. Clearly, the positions $k+1$ to $n$ of $H$ are free of stopping sets of size $s \leq n-k$. Since the automorphism group of $\mathcal{C}$ is $s$-transitive, any collection of not more than $s$ coordinates in $[0, n-1]$ is moved by some element of $\text{Aut}(\mathcal{C})$ into the parity-positions. Consequently, the automorphism group itself represents a (possibly non-minimal) **s-SAD** set. ∎

Finding **SAD** sets of codes is a very complicated task, so that we focus our attention on deriving bounds on the size of such sets for specific examples of codes. In this direction, we have the following upper bound for $\mathbf{S}^\star$ of the $[24, 12, 8]$ Golay code. First, since the automorphism group of the Golay code is 5-fold transitive, **5-SAD** sets exist, and $\mathbf{S}_5^\star \leq |\mathcal{M}_{24}|$, where $|\mathcal{M}_{24}|$ denotes the order of the Mathieu group $\mathcal{M}_{24}$, which equals 244823040. Our next result shows that $\mathbf{S}^\star$ is actually much smaller.

*Theorem 3.3:* Let $\mathcal{G}_{24}$ be the unique $[24, 12, 8]$ Golay code. Then $\mathbf{S}_5^\star(\mathcal{G}_{24}) \leq 14$.

*Proof:* The proof is constructive and based on a collection of results regarding the stopping redundancy hierarchy of the Golay code [5] and **PD** sets [10]. Due to space limitations, we omit the details of the proof. Below, we list the particular form of $H$ used to meet the claimed results, as well as the set of corresponding automorphisms. The matrix in question is $H(\mathcal{G}_{24}) = [I_{12}|M]$, where

$$M = \begin{bmatrix} I_3 & A & A^2 & A^4 \\ A & I_3 & A^4 & A^2 \\ A^2 & A^4 & I_3 & A \\ A^4 & A^2 & A & I_3 \end{bmatrix}, \quad A = \begin{bmatrix} 1 & 1 & 1 \\ 1 & 0 & 0 \\ 1 & 0 & 1 \end{bmatrix}. \tag{3}$$

The automorphisms are of the form $\theta^i \times \pi^j$, $i = 0, 1$, $j = 0, 1, \ldots, 5, 6$, where

$$\theta = (0, 12)(1, 13)(2, 14)(3, 15) \ldots (10, 22)(11, 23);$$
$$\pi = (3, 6, 15, 9, 21, 18, 12)(4, 7, 16, 10, 22, 19, 13) \tag{4}$$
$$(5, 8, 17, 11, 23, 20, 14);$$

in the standard cycle form. ∎

The cardinality of a **PD** set for the Golay code is also known, and a **PD** set can be shown to consist of the same set of 14 permutations described in the above theorem (see [10]). This leads to a bound on $\rho_5(\mathcal{G}_{24}) \leq 14 \times 12 = 168$ which is significantly larger than the constructive bound of value 24 found by the authors in [5], but also significantly better than the general bound obtained from Lovasz Local Lemma described in [5]. The parity-check matrix meeting this bound will be denoted by $H_W$ as Eq. (3) and Eq. (4) were introduced in [10].

For the Golay code, one can show an even stronger result. Assume that the underlying parity-check matrix of $\mathcal{G}_{24}$ is of the form below, denoted by $H^\star$, and that a set of 23 automorphisms of the form $\epsilon, \tau, \tau^2, \ldots, \tau^{22}$, where $\epsilon$ denotes the identity element of $S_{24}$ (the symmetric group of order 24), and $\tau = (0\ 1\ 2\ \ldots\ 21\ 22)(23)$, is used. Then up to weight 11, all undecodable erasure patterns[2] for an automorphism group decoder of $\mathcal{G}_{24}$ are exactly the undecodable erasure patterns of a maximum likelihood (ML) decoder of the code. The number of undecodable erasure patterns of each weight for the described decoder and a ML decoder are shown in Table I.

$$\boldsymbol{H}^\star = \begin{pmatrix} 1\,1\,1\,0\,0\,0\,0\,0\,1\,0\,0\,1\,1\,0\,0\,0\,0\,0\,1\,0\,0\,0\,0\,1 \\ 1\,1\,0\,0\,0\,0\,1\,0\,0\,0\,0\,1\,0\,0\,1\,1\,1\,0\,0\,0\,0\,0\,0\,1 \\ 1\,1\,0\,1\,0\,0\,1\,0\,1\,0\,1\,0\,0\,1\,0\,0\,0\,0\,0\,0\,0\,0\,0\,1 \\ 1\,1\,1\,0\,0\,0\,1\,1\,0\,0\,0\,0\,0\,0\,0\,0\,0\,0\,0\,1\,0\,1\,0\,1 \\ 1\,1\,0\,0\,0\,1\,0\,0\,0\,1\,0\,1\,0\,1\,0\,0\,0\,0\,0\,0\,0\,1\,0\,1 \\ 1\,1\,0\,1\,0\,0\,0\,1\,0\,1\,0\,0\,0\,0\,0\,0\,1\,0\,1\,0\,0\,0\,0\,1 \\ 0\,1\,1\,0\,0\,0\,1\,0\,0\,0\,0\,1\,0\,1\,0\,0\,0\,1\,0\,0\,1\,0\,0\,1 \\ 1\,1\,0\,1\,1\,0\,0\,0\,0\,0\,0\,1\,0\,0\,0\,0\,0\,0\,0\,0\,1\,1\,0\,0\,1 \\ 1\,1\,1\,1\,0\,1\,0\,0\,0\,0\,0\,0\,0\,0\,1\,0\,0\,1\,0\,0\,0\,0\,0\,1 \\ 1\,1\,0\,0\,1\,0\,0\,0\,0\,0\,1\,0\,0\,0\,1\,0\,0\,0\,1\,0\,0\,1\,0\,1 \\ 0\,0\,1\,1\,0\,0\,1\,0\,1\,0\,0\,1\,0\,0\,1\,0\,0\,0\,0\,1\,0\,0\,0\,1 \\ 0\,0\,1\,1\,0\,1\,0\,1\,0\,0\,1\,1\,0\,0\,1\,0\,0\,0\,0\,0\,0\,0\,1\,0 \end{pmatrix}$$

We conclude this brief overview by pointing out that there exists a strong connection between the problem of set coverings [11] and the problem of finding **PD** and **SAD** sets. This relationship can be used to develop simple heuristic search strategies for parity-check matrices amenable for automorphism group decoding. The crux of the approach lies in identifying subsets of columns of $\boldsymbol{H}$ that have cardinality larger than $n-k$ and that are free of stopping sets of size up to and including $d-1$, and in combining these results with ideas borrowed from set covering theory. More details regarding this strategy will be given in the full version of the paper.

IV. AUTOMORPHISM GROUP DECODERS FOR THE BEC

In this section, we study the performance of automorphism group decoders (AGD) on cyclic and extended cyclic codes. We restrict our attention to these classes of codes since large subgroups of the automorphism group of such codes are known and since the implementation complexity of AGD decoders in this case is very small. Nevertheless, the described decoding techniques can be applied to other classes of codes for which some information about the automorphism group is available. For the case of extended cyclic codes, it is tacitly assumed that the overall parity-check bit is confined to the last position of the codewords and that its index is $n-1$.

Throughout the section, we make use of the following result.

*Theorem 4.1:* [7] Let $\boldsymbol{c} = (c_0, c_1, \ldots, c_{n-1})$ be a codeword of an $[n,k,d]$ cyclic code. The automorphism group of the code contains the following two sets of permutations, denoted by $C_1$ and $C_2$.

$C_1$: Cyclic permutations $\gamma, \gamma^2, \ldots, \gamma^{n-1}$:
  $\gamma\,\boldsymbol{c} = (c_{n-1}, c_0, c_1, \ldots, c_{n-2})$;
$C_2$: Permutations $\zeta, \zeta^2, \ldots, \zeta^m$, where $\zeta : i \to 2 \cdot i \bmod n$, and where $m$ denotes the cardinality of the cyclotomic coset (of the $n$-th roots of unity) that contains one:

[2]I.e., erasure patterns that contain the support of a codeword of $\mathcal{G}_{24}$.

  $\zeta\,\boldsymbol{c} = (c_0\,c_2\,\ldots\,c_{2\,(n-1)})$. Note that all subscripts in the vectors described above are taken modulo $n$.

For extended cyclic codes, we use the same notation $C_1$ and $C_2$ to describe permutations that fix $c_{n-1}$ and act on the remaining coordinates as described in the theorem above. All automorphisms can be decomposed into products of disjoint cycles. Permutations in $C_1$ have one single cycle (or two cycles, for the case of extended cyclic codes), while permutations in $C_2$ have a number of cycles that equals the number of cyclotomic cosets $r$ of the $n$-roots of unity (or $r+1$ cycles, for the case of extended cyclic codes).

*Definition 4.1:* We say that a parity-check matrix $\boldsymbol{H}$ *resolves* a set of coordinates if those coordinates do not correspond to a stopping set in $\boldsymbol{H}$.

In order to describe the parity-check matrices used for automorphism group decoding of cyclic codes, we introduce the notion of the *cyclic orbit generator* [5].

*Definition 4.2:* Let $\mathcal{C}$ be an $[n,k,d]$ binary linear cyclic code. Partition the set of codewords in the dual code $\mathcal{C}^\perp$ of minimum weight into sets consisting of cyclic shifts of one given codeword. Alternatively, partition the set of minimum weight codewords into orbits of the cyclic group. We refer to one chosen representative from each orbit as the *cyclic orbit generator* (cog).

AGD$_A$ **Decoders**: These decoders only employ permutations from the set $C_1$, which reduces the permutation architecture to one shift register. The decoder uses a standard iterative decoding algorithm until the presence of a stopping set is detected. In that case, AGD$_A$ applies a randomly chosen cyclic shift to the current word. If the iterative decoder encounters another stopping set, the whole process is repeated with a (different) cyclic permutation. The decoding process terminates if either all permutations in $C_1$ are tested or if the decoder successfully recovers the codeword.

Assume that the number of cogs of a cyclic code is at least $n-k$. In this case, the parity-check matrix used for decoding consists of $n-k$ different cogs, provided that such a matrix has full rank. A redundant parity-check matrix consisting of the collection of all vectors in the orbits of the cogs will be denoted by $\boldsymbol{H}_{\text{AGD}_A}$.

AGD$_B$ **Decoders**: These decoders use permutations drawn from both $C_1$ and $C_2$. If a stopping set is encountered, the decoder first tries to resolve this set by applying a randomly chosen permutation from $C_1$. Only after the whole set $C_1$ is exhausted, a permutation from $C_2$ is applied to the current decoder word. Before presenting simulation results for two particular codes, the $[24, 12, 8]$ Golay code and the cyclic $[31, 16, 7]$ BCH code, we formally introduce the notion of undecodable erasure patterns.

*Definition 4.3:* An undecodable erasure pattern of size $\sigma$ is either a stopping set of size $\sigma$ or a set that properly contains a stopping set.

A. $[24, 12, 8]$ *Golay code*

In Table I we list the number of undecodable erasure patterns of size up to $\sigma = 12$ encountered in several parity-

TABLE I

UNDECODABLE ERASURE PATTERNS OF THE $[24, 12, 8]$ GOLAY CODE.

| $\sigma$ | Number of undecodable erasure patterns | | | | |
|---|---|---|---|---|---|
| | $\boldsymbol{H}^{\star}$ | $\boldsymbol{H}_{\mathrm{AGD}_A}$ | $\boldsymbol{H}_{\mathrm{W}}$ | [4] | ML |
| 3 | 7 | 0 | 0 | 0 | 0 |
| 4 | 190 | 0 | 0 | 0 | 0 |
| 5 | 2231 | 0 | 0 | 0 | 0 |
| 6 | 15881 | 0 | 0 | 0 | 0 |
| 7 | 79381 | 0 | 0 | 0 | 0 |
| 8 | 293703 | 759 | 759 | 3284 | 759 |
| 9 | 805556 | 12144 | 12158 | 78218 | 12144 |
| 10 | 1613613 | 91080 | 93477 | 580166 | 91080 |
| 11 | 2378038 | 425040 | 481764 | 1734967 | 425040 |
| 12 | 2690112 | 1322178 | $\geq 481764$ | $\geq 1734967$ | 1313116 |
| $\geq 13$ | $\binom{24}{\sigma}$ | $\binom{24}{\sigma}$ | $\binom{24}{\sigma}$ | $\binom{24}{\sigma}$ | $\binom{24}{\sigma}$ |

check matrices of the Golay $[24, 12, 8]$ code; $\boldsymbol{H}^{\star}$ and $\boldsymbol{H}_{\mathrm{W}}$ are specified in Section III, while the matrix referenced by [4] corresponds to a matrix of dimension $34 \times 24$ described in the given paper. ML refers to the erasure patterns that cannot be recovered by a ML decoder, while $\boldsymbol{H}_{\mathrm{AGD}_A}$ corresponds to the matrix described in this section. Figure 1 shows the performance of iterative decoders operating on the parity-check matrices in Table I. As can be noted, there is a significant performance gain of $\mathrm{AGD}_A$ or $\mathrm{AGD}_B$ decoders when compared to standard belief-propagation decoders operating on the redundant parity-check matrix described in [4] or on any other form of standard Tanner graphs. For $\mathrm{ER} \leq 0.15$, all matrix representations require an almost equal number of iterations, indicated by the vertical bar in Figure 1.

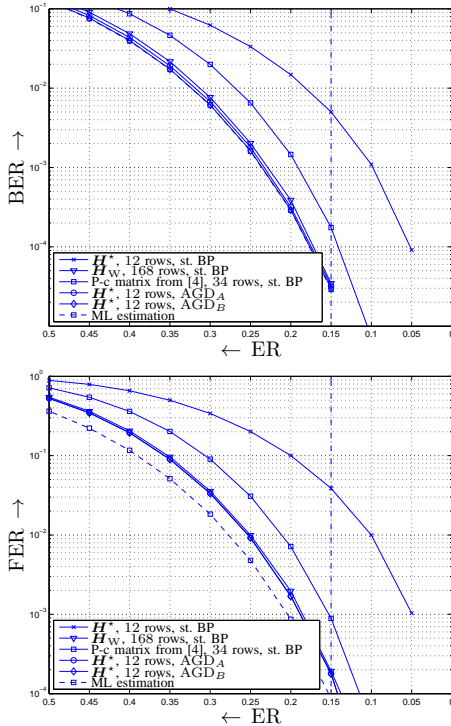

Fig. 1. Performance comparison for the $[24, 12, 8]$-Golay code.

### B. $[31, 16, 7]$ BCH code

The dual of the $[31, 16, 7]$ BCH code contains 465 codewords of weight 7. These codewords can be partitioned into 15 cyclic orbits. Figure 2 shows the performance of various types of iterative decoders for the given BCH code. The $\mathrm{AGD}_A$ decoder operated on a parity-check matrix that consists of all 15 cogs, and for comparison, standard iterative decoding is performed on the same matrix. In addition, we show the performance of standard iterative decoders on three different cyclic parity-check matrices with $m = n - k = 15$, $m = 21$, and $m = 31$ rows. The generator codeword for these matrices is $\mathrm{cog}_{31,16,A} = [1\,4\,1\,4\,0\,5\,0\,0\,0\,2\,2]$, given in octal form and with the most significant bit on the left hand side.

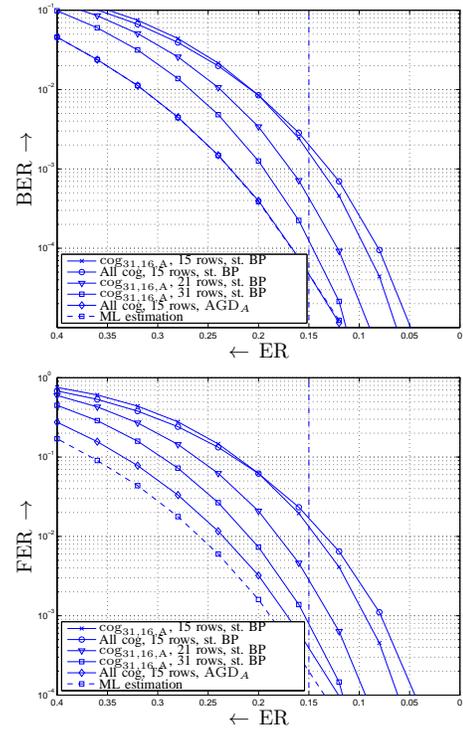

Fig. 2. Performance comparison for the $[31, 16, 7]$-BCH code.